\documentclass[12pt]{article}
\usepackage{setspace}
\usepackage[a4paper]{geometry}
\usepackage{amsmath}
\usepackage{amssymb}
\usepackage{authblk}
\usepackage{float}
\usepackage{graphicx}
\graphicspath{{figures/}}
\usepackage[hidelinks, colorlinks = TRUE,
            linkcolor=red,
            urlcolor=blue,
            citecolor=gray]{hyperref}
\usepackage{natbib}
\usepackage{subcaption}
\usepackage{xcolor}

\usepackage{chngcntr}

\title{Drivers of Success: A Bayesian State-Space Model to Disentangling Latent Driver and Constructor Abilities in Formula One}

\author[1]{Tim Lindner}
\author[1, 2]{Rui Jorge Almeida}
\author[1]{Nalan Ba\c{s}t\"{u}rk}
\author[1]{Stephan Smeekes}
\affil[1]{Department of Quantitative Economics, Maastricht University}
\affil[2]{Department of Data Analytics and Digitalisation, Maastricht University}

\date{\today}

\begin{document}

\maketitle

\begin{abstract}
Formula One outcomes reflect the joint contributions of drivers and constructors, but these contributions are unobserved and vary over time. We propose a Bayesian state-space model that disentangles dynamic driver and constructor abilities using two observed outcomes: fastest qualifying lap times and race rankings. Both outcomes depend jointly on latent driver and constructor states that evolve at the Grand Prix level, while the race equation additionally accounts for starting-grid position. The decomposition is supported by constraints that center the driver and constructor abilities at zero, together with variation in driver--constructor assignments over time. Bayesian inference is performed using the No-U-Turn sampler under weakly informative priors that treat driver and constructor abilities symmetrically. Applying the model to the Formula One hybrid era from 2014 to 2021, we find substantial heterogeneity in both driver and constructor abilities. Driver abilities are generally more stable over time, whereas constructor abilities exhibit greater variation and, for many driver--constructor combinations, contribute more strongly to observed performance.
\end{abstract}



\onehalfspacing

%
%

\section{Introduction}

Formula One performance is jointly determined by a driver and the constructor for which the driver competes. Qualifying times and race rankings are observed, but the respective contributions of driver and constructor ability are not. Separating these contributions is complicated by their time-varying nature, changes in driver--constructor assignments, and the different measurement scales of the observed outcomes. We develop a Bayesian state-space model that jointly analyzes these outcomes and decomposes their shared dynamic predictor into driver and constructor abilities.

The qualifying and race outcomes are heterogeneous outcome types, by which we mean that they are defined on different measurement scales and modeled using outcome-specific sampling distributions. A major methodological contribution of this paper is to model these two outcomes jointly. Mixed-outcome models commonly connect continuous, binary, ordinal, or count responses through shared latent variables, allowing information to be combined across outcomes while retaining an appropriate sampling distribution for each response \citep{MoustakiKnott2000,Dunson2000,GueorguievaAgresti2001,Bhat2015}. Most foundational models in this literature are static, although longitudinal extensions allow serial dependence between repeated mixed outcomes \citep[for example,][]{SeedorffEtAl2023}. Bayesian state-space and dynamic generalized linear models separately provide general frameworks in which latent states or regression coefficients evolve over time \citep{WestHarrisonMigon1985,Gamerman1998,chan2023bayesian}. The proposed model brings these ideas together by linking a continuous outcome and a ranking outcome through the same dynamic latent states.

A related literature develops dynamic models for paired-comparison and ranking outcomes. Dynamic extensions of the Bradley-Terry and Plackett-Luce models allow latent worth parameters to evolve through exponential weighting, stochastic time-series processes, or autoregressive latent scores \citep{CattelanVarinFirth2013,HendersonKirrane2018,HolyZouhar2022,IacopiniONeillRossini2026}. These models accommodate repeated paired comparisons, permutations, or partial rankings, but their latent structure is generally organized around the evolving worth of the ranked entities. In Formula One, \cite{HendersonKirrane2018} use time-weighted Plackett-Luce models to forecast race results, while \cite{Yamauchi2021} estimates dynamic driver and constructor ratings from race rankings in separate models. These approaches do not use a continuous outcome jointly with the ranking outcome to estimate an additive decomposition of shared performance.

The structured decomposition of performance into driver and constructor components is closely related to two-way worker-firm models in econometrics. Beginning with \cite{AbowdKramarzMargolis1999}, this literature decomposes wages into substantively labelled worker and firm effects and uses worker mobility across firms to separate their contributions. Correlated-random-effects and empirical-Bayes approaches additionally use prior structure and the observed bipartite assignment graph to improve the estimation of two-way effects under limited mobility \citep{BonhommeEtAl2023,ChengHoSchorfheide2025}. Time-varying extensions allow firm effects to change over time \citep{EngbomMoserSauermann2023}. These models provide an important analogue to the driver--constructor assignment structure, but they generally analyze a single continuous outcome and provide fewer examples in which both labelled effect systems follow separate stochastic evolution equations within a joint state-space model.

Some contributions in Formula One statistics also separate driver and constructor effects \citep{Eichenberger2009,Bell2016,Kesteren2023}. These models generally focus on a single observed outcome or impose more limited temporal structures. The present paper connects this decomposition problem with the mixed-outcome, dynamic-ranking, and state-space literatures.

In this paper, we propose a Bayesian state-space model for two Formula One outcomes: standardized fastest qualifying lap times and race finishing positions. Qualifying time is a continuous outcome, whereas a race result is a partial ranking outcome, represented by a permutation of the drivers who finish the race. The two outcomes jointly depend on a shared dynamic predictor that is decomposed additively into driver and constructor abilities. Both sets of abilities evolve at the Grand Prix level, and the observed contractual assignment determines which constructor ability enters the performance of each driver. Given that qualifying outcomes determine the starting-grid of a race, we additionally include grid position in the race equation.

Identification of the dynamic decomposition is an essential part of the model. The continuous qualifying equation and the ranking likelihood provide complementary information about the shared latent predictor. Separate sum-to-zero constraints are imposed on the driver and constructor abilities, so that each set of abilities is centered at zero at every Grand Prix. Variation in observed driver--constructor assignments supplies the information needed to separate the two components, provided that the assignment structure connects the constructors through driver changes. This combination of likelihood information, parameter restrictions, prior specification, and observed assignment design identifies the decomposition. The dependence of identification on comparison and assignment structures is also central in two-way-effects and covariate-assisted ranking models \citep{ChengHoSchorfheide2025,DongEtAl2026}.

The model is tailored to the characteristics of Formula One competition. Qualifying times measure one-lap performance, whereas race rankings reflect sustained performance under evolving conditions. We model fastest qualifying lap times to obtain a granular measure of one-lap performance and finishing positions because race finishing times can be distorted by pace management and other strategic considerations. We use the same driver and constructor states in both equations because the objective is to estimate holistic abilities shared across qualifying and race performance, rather than outcome-specific ability measures.

Bayesian inference for the Formula One hybrid era from 2014 to 2021 is performed using the No-U-Turn sampler under weakly informative priors that treat driver and constructor states symmetrically. The posterior estimates indicate that driver abilities are generally more stable than constructor abilities over the analyzed period. Constructor abilities exhibit greater variation and, for several driver--constructor combinations, make a larger relative contribution to the shared performance predictor.

The remainder of this paper is structured as follows. Section~\ref{sec_model} introduces the Formula One outcomes and presents the dynamic latent-variable model. Section~\ref{sec_identification_priors} discusses identification, prior specification, and posterior inference. Section~\ref{sec_case-study} applies the model to the Formula One hybrid era, and Section~\ref{sec_conclusion} concludes.

\section{A Dynamic driver--constructor Decomposition Model}
\label{sec_model}

Formula One Grand Prix weekends consist of a qualifying session and a race. Qualifying performance determines the starting-grid for the race, and each driver competes using a car developed and operated by the constructor with which the driver is contracted. Formula One constructors represent distinct sources of performance variation because they develop their own cars and differ in their technical, operational, and financial resources \citep{Kesteren2023}. driver--constructor assignments may also change over time.

The proposed model combines two heterogeneous outcome types: a continuous qualifying outcome and a partial ranking race outcome. Each outcome is modeled using an outcome-specific sampling distribution, while both depend on the same evolving latent driver and constructor abilities. Mixed-outcome latent-variable models similarly connect outcomes defined on different measurement scales through shared unobserved quantities \citep{MoustakiKnott2000,Dunson2000,GueorguievaAgresti2001,Bhat2015}. In the present setting, this principle is extended dynamically: both outcomes depend on a shared latent performance variable that evolves over time. This variable is then decomposed into separate dynamic driver and constructor abilities.


We consider two observed outcomes: the qualifying time $q_{d,t}$ and race rank $r_{d,t}$ of driver $d=1,\ldots,D$ at Grand Prix $t=1,\ldots,T$. Their observation equations are:
\begin{align}
\label{eq_model_core_dgp_quali-time}
\text{if } i_{d,t}^{\left(d\right)} = 1&: \quad q_{d,t} = \mu_{d,t}^{\left(q\right)} + \epsilon_{d,t}^{\left(q\right)}
; \quad \epsilon_{d, t}^{\left(q\right)} \overset{\mathrm{iid}}{\sim} \mathcal{N}\left(0, \sigma^{\left(q\right)}\right), \\
\label{eq_model_core_dgp_race-rank}
\text{if } f_{d,t}^{\left(r\right)} = 1&: \quad r_{d,t} = \sum_{u = 1}^{U_t}u \times \boldsymbol{1} \left(p_{d, t} = p_{\left(u\right) t}\right)
; p_{d,t} = \mu_{d,t}^{\left(r\right)} + \epsilon_{d,t}^{\left(r\right)}
; \quad \epsilon_{d, t}^{\left(r\right)} \overset{\mathrm{iid}}{\sim} \mathcal{G}\left(0, \sigma^{\left(r\right)}\right), 
\end{align}
where $\mu_{d,t}^{(q)}$ and $\mu_{d,t}^{(r)}$ denote the outcome-specific performance predictors for qualifying and the race, respectively. The terms $\epsilon_{d,t}^{(q)}$ and $\epsilon_{d,t}^{(r)}$ are outcome-specific disturbances, and $i_{d,t}^{\left(d\right)}=1$ indicates that driver $d$ participates in Grand Prix $t$, while $f_{d,t}^{\left(r\right)}=1$ indicates that the driver finishes the race. Thus, the qualifying equation is defined for participating drivers, whereas the race-ranking equation is defined for race finishers. Furthermore, the qualifying time is constructed from the fastest lap to measure the driver's one-lap performance potential. 

In equations~\eqref{eq_model_core_dgp_quali-time} and \eqref{eq_model_core_dgp_race-rank}, superscripts distinguish the four indicators used in the model: $i_{d,t}^{\left(d\right)}$ and $i_{k,t}^{\left(k\right)}$ indicate whether driver $d$ participates and whether constructor $k$ is represented in Grand Prix $t$, respectively. On the other hand, $f_{d,t}^{\left(q\right)}$ and $f_{d,t}^{\left(r\right)}$ indicate whether driver $d$ records a qualifying time and finishes the race, respectively. The dynamics of abilities for participating and non-participating drivers and for represented and non-represented constructors are introduced separately in the state equations.

Let $\widetilde{q}_{d,t}$ denote the raw qualifying time. For drivers who record a valid qualifying lap, $\widetilde{q}_{d,t}$ is the fastest lap time achieved across Q1, Q2, and Q3. A driver may participate in qualifying without recording a valid lap time, for example because of a crash or mechanical problem. In this case, the driver's raw qualifying time is imputed at $107$ percent of the fastest recorded time:
\begin{equation}
\text{if } i_{d,t}^{\left(d\right)} = 1 \text{ and } f_{d,t}^{\left(q\right)} = 0: \quad \widetilde{q}_{d,t}= 1.07 \,
 \operatorname{min}\left(\widetilde{\boldsymbol{q}}_t\right),
\end{equation}
where $\widetilde{\boldsymbol{q}}_t$ is the vector of qualifying lap times for participating drivers at Grand Prix $t$ and $\text{min}(\cdot)$ denotes the minimum computed after excluding missing values. This treatment is motivated by the Formula One 107 percent rule, under which a driver whose best Q1 lap exceeds 107 percent of the fastest Q1 time, or who fails to set a time, may not start the race unless admitted by the stewards under exceptional circumstances \citep[Art.~36.2]{FIA2021SportingRegulations}.

Raw qualifying lap times are not directly comparable across Grand Prix weekends because circuits, weather conditions, and other event-specific circumstances differ. We therefore transform the lap times into within-qualifying standard scores:

\begin{equation}
\text{if } i_{d,t}^{\left(d\right)} = 1: \quad
q_{d,t} =
\frac{\widetilde{q}_{d,t}-\bar{\widetilde{q}}_t}
{s_{\widetilde{q},t}},
\end{equation}
where $\bar{\widetilde{q}}_t$ and $s_{\widetilde{q},t}$ denote the mean and standard deviation, respectively, of the observed and imputed raw qualifying times at Grand Prix $t$. We subsequently refer to the standardized outcome $q_{d,t}$ as the qualifying time, which is the outcome variable in equation~\eqref{eq_model_core_dgp_quali-time}.

The race outcome is the official partial ranking of the drivers who finish race $t$. The observed race rank $r_{d,t}$ denotes the position assigned to driver $d$ within this ranking. We note that the observed race rank can be different than the driver's on-track finishing position because penalties or disqualifications may alter the official race classification.

We exclude drivers who do not finish the race from the ranking likelihood. The indicator $f_{d,t}^{\left(r\right)}$ equals one if driver $d$ finishes race $t$ and zero otherwise. The race model therefore describes the ranking of drivers conditional on finishing. The number of drivers included in the partial ranking at race $t$ is:
\begin{equation}
\label{eq:definition_Ut}
U_t = \sum_{d=1}^D f_{d,t}^{\left(r\right)}.
\end{equation}

The qualifying outcome is thus a continuous scalar, while the race outcome is a partial ranking represented by a permutation of the drivers included in the race likelihood.
Equations~\eqref{eq_model_core_dgp_quali-time}-\eqref{eq_model_core_dgp_race-rank} connect the two observed outcomes to outcome-specific latent performance predictors. The qualifying time depends on $\mu_{d,t}^{(q)}$, while the race rank is induced by the latent race performance $p_{d,t}$ with systematic component $\mu_{d,t}^{(r)}$.
The indicator function $\boldsymbol{1}(\cdot)$ takes the value of one when its argument is true and zero otherwise, and $p_{\left(u\right)t}$ denotes the $u$th order statistic of the latent race performances after excluding missing values. Under the sign convention used here, a lower value of either latent performance predictor corresponds to a better expected outcome.

We use a Normal distribution for standardized qualifying times. This provides a parsimonious model for a continuous outcome recorded under the comparatively controlled conditions of a qualifying session. The race outcome requires a distribution over rankings rather than independent distributions for the individual observed ranks. Under the Gumbel specification for latent race performances, the race likelihood can be represented analytically using a Plackett-Luce model, as developed in Section~\ref{sec_identification_priors} \citep{Glickman2015}. Dynamic Bradley-Terry and Plackett-Luce models similarly represent changing competitive performance through the time-variation in latent states \citep{CattelanVarinFirth2013,HendersonKirrane2018,HolyZouhar2022,IacopiniONeillRossini2026}.

The Normal and Gumbel specifications therefore perform different roles: the former defines the distribution of the continuous qualifying outcome, and the latter induces a probability distribution over the observed race ranking. The model does not require the heterogeneous outcomes to share a sampling distribution; instead, they are linked through the shared latent components.


Both outcome-specific performance predictors share a common additive decomposition into driver and constructor abilities. Let $a_{d,t}$ denote the ability of driver $d$ at Grand Prix $t$, and let $c_{k,t}$ denote the ability of constructor $k=1,\ldots,K$. The qualifying and race performance predictors are:
\begin{align}
\label{eq_model_core_dgp_quali-time_mean_err}
\text{if } i_{d,t}^{\left(d\right)} = 1&: \quad \mu_{d,t}^{\left(q\right)} = a_{d,t} + \sum_{k=1}^K c_{k,t} \,\boldsymbol{1}(z_{d,t} = k), \\
\label{eq_model_core_dgp_race-performance_mean_err}
\text{if } f_{d,t}^{\left(r\right)} = 1&: \quad \mu_{d,t}^{\left(r\right)}
= \beta_1 \left(a_{d,t} + \sum_{k=1}^K c_{k,t} \, \boldsymbol{1}(z_{d,t} = k)\right)
+ \beta_2 g_{d,t},
\end{align}
where $z_{d,t}\in{1,\ldots,K}$ identifies the constructor with which driver $d$ is contracted at Grand Prix $t$.

The constructor component entering a driver's performance predictor is therefore determined by an observed assignment. When a driver changes constructors, the assignment variable $z_{d,t}$ changes accordingly. Additive decompositions into individual and organizational components connected through observed assignments are familiar from two-way worker-firm models \citep{AbowdKramarzMargolis1999,BonhommeEtAl2023,ChengHoSchorfheide2025}. The proposed model differs by allowing both component systems to evolve over time and by using them jointly to generate a continuous outcome and a partial ranking outcome.

The shared driver--constructor component enters the qualifying predictor directly. In the race equation, its contribution is scaled by $\beta_1$, while $\beta_2$ captures the association between the driver's starting-grid position $g_{d,t}$ and latent race performance. Because the grid is determined by qualifying, the race equation describes the contribution of driver and constructor abilities to the race ranking conditional on starting position. The coefficients $\beta_1$ and $\beta_2$ jointly determine the relative contributions of the shared ability component and grid position to latent race performance.

We impose the same driver and constructor states in the qualifying and race equations rather than introducing outcome-specific abilities. This is an important restriction: driver and constructor abilities are interpreted as shared characteristics that affect both one-lap qualifying performance and sustained race performance, although the two outcomes have different observation equations.

Conditional on the driver and constructor abilities and, for the race outcome, grid position, we assume that the qualifying and race errors are independent. The errors represent the remaining outcome-specific variation not explained by the modeled latent states and observed grid position.


Driver and constructor abilities vary over Grand Prix weekends. We model their evolution using separate random-walk state equations:
\begin{align}
\label{eq_model_core_dgp_gp-ability_rw1}
\begin{split}
\text{if } i_{d,t}^{\left(d\right)} = 1: \quad a_{d,t} &= a_{d,t-1} + \epsilon_{d,t}^{\left(a\right)};
\quad \epsilon_{d,t}^{\left(a\right)}\overset{\mathrm{iid}}{\sim} N(0,\sigma^{(a)});\\
\text{if } i_{d,t}^{\left(d\right)} = 0: \quad a_{d,t} &= a_{d,t-1},
\end{split}\\
\label{eq_model_core_dgp_gp-ability_rw2}
\begin{split}
\text{if } i_{k,t}^{\left(k\right)} = 1: \quad c_{k,t} &= c_{k,t-1} + \epsilon_{k,t}^{\left(c\right)};
\quad \epsilon_{k,t}^{\left(c\right)}\overset{\mathrm{iid}}{\sim} N(0, \sigma^{(c)});\\
\text{if } i_{k,t}^{\left(k\right)} = 0: \quad c_{k,t} &= c_{k,t-1},
\end{split}\\
\label{eq:model_core_dgp-ability-init-drv}
a_{d,0} &\overset{\mathrm{iid}}{\sim} \mathcal{N} \left(\bar{a}, \varsigma^{\left(a\right)}\right),\\
\label{eq:model_core_dgp-ability-init-ctr}
c_{k,0} &\overset{\mathrm{iid}}{\sim} \mathcal{N} \left(\bar{c}, \varsigma^{\left(c\right)}\right),
\end{align}
where the ability of driver $d$ remains constant when the driver does not participate in Grand Prix $t$. The same mechanism applies to a constructor that does not participate.

Random-walk dynamics allow driver and constructor abilities to change gradually without imposing mean reversion or a deterministic time profile. Stochastic state evolution is standard in Bayesian dynamic generalized linear and state-space models \citep{WestHarrisonMigon1985,Gamerman1998,chan2023bayesian}. Existing dynamic ranking models also allow entity-specific latent states to vary over time, but generally evolve a single class of states associated with the ranked entities. Here, driver and constructor abilities form two separately evolving states linked by the observed assignment structure.
The assumptions of uncorrelated innovations in the state equations can be relaxed. In Section~\ref{sec_identification_priors}, we introduce correlated driver innovations and correlated constructor innovations that impose the restrictions required for the dynamic decomposition.

Equations~\eqref{eq_model_core_dgp_quali-time}-\eqref{eq:model_core_dgp-ability-init-ctr} jointly define the proposed state-space model. The observation equations combine a continuous qualifying outcome and a partial ranking race outcome using outcome-specific sampling distributions. Both outcomes depend on shared driver and constructor states, while the observed assignment $z_{d,t}$ determines which constructor state enters each driver's predictor. The two state variables evolve separately over Grand Prix weekends. Relative to static mixed-outcome models, the shared latent components are dynamic; relative to dynamic ranking models, the ranking is analyzed jointly with a continuous outcome. In many latent-factor models, the relationship between latent factors and observed outcomes is estimated through factor loadings. In the proposed model, the relevant driver and constructor components are instead determined by the observed driver--constructor assignments. The identification restrictions required to separate these components are developed in the following section.

\section{Identification and Bayesian Inference}
\label{sec_identification_priors}

Identification in the proposed model involves several related requirements. Dynamic latent-state models require restrictions that determine the locations and scales of the states. The partial ranking likelihood introduces additional invariances because rankings depend on relative rather than absolute latent performances. Finally, the additive decomposition of the shared predictor into separately evolving driver and constructor abilities creates a further identification problem. We address these requirements using the joint-outcome likelihood, parameter restrictions implemented through the prior specification, and variation in observed driver--constructor assignments.


The model in equations~\eqref{eq_model_core_dgp_quali-time}--\eqref{eq:model_core_dgp-ability-init-ctr} implies the following decomposition of the joint log likelihood:
\begin{equation}
    \label{eq_likelihood}
    \begin{split}
        \ell\left(\boldsymbol{Q}, \boldsymbol{R} \, | \, \boldsymbol{A}, \boldsymbol{C}, \boldsymbol{I}^{\left(d\right)}, \sigma^{\left(q\right)}, \boldsymbol{F}^{\left(r\right)}, \boldsymbol{\beta}, \sigma^{\left(r\right)}\right)
        &= \left(\sum_{t=1}^T \sum_{d=1}^D \ell\left(q_{d,t} \, | \, i_{d,t}^{\left(d\right)}, \mu_{d,t}^{\left(q\right)}, \sigma^{\left(q\right)}\right)\right) \\
        &\quad+ \sum_{t=1}^T \ell\left(\boldsymbol{r}_t \, | \, \boldsymbol{f}^{\left(r\right)}, \boldsymbol{\mu}_t^{\left(r\right)}, \sigma^{\left(r\right)}\right),
    \end{split}
\end{equation}
where $\boldsymbol{Q}= \{q_{1,1}, \ldots, q_{D, T}\}$ and $\boldsymbol{R}= \{\boldsymbol{r}_1, \ldots, \boldsymbol{r}_T\}$ are the set of all observed outcomes;
    $\boldsymbol{I}^{(d)}= \{i^{(d)}_{1,1}, \ldots, i^{(d)}_{D, T}\}$ and   $\boldsymbol{F}^{(r)}= \{\boldsymbol{f}_1, \ldots, \boldsymbol{f}^{(r)}_T\}$ are the set of assignment indicators;
and $\boldsymbol{A} = \{a_{1,1}\ldots, a_{D,T}\}$ and $\boldsymbol{C} = \{c_{1,1}\ldots, c_{K,T}\}$ are the set of latent state variables.

The first part in~\eqref{eq_likelihood} corresponds to the qualifying observation equation and can be written in more detail as:
\begin{equation}
    \begin{split}
        \text{if } i_{d,t}^{\left(d\right)} = 1&: \quad \ell\left(q_{d,t} \, | \, \mu_{d,t}^{\left(q\right)}, \sigma^{\left(q\right)}\right) = \log \mathcal{N}\left(\mu_{d,t}^{\left(q\right)}, \sigma_{\left(q\right)}\right); \\
        \text{if } i_{d,t}^{\left(d\right)} = 0&: \quad \ell\left(q_{d,t}\right) = 0.
    \end{split}
\end{equation}
The second part of the joint log likelihood corresponds to the race observation equation. The two likelihood contributions use outcome-specific sampling distributions but depend on the same evolving driver and constructor abilities. The joint-outcome structure therefore allows both the continuous qualifying outcomes and the partial race rankings to inform the shared latent states.

The Gumbel distribution in equation~\eqref{eq_model_core_dgp_race-rank} also provides a convenient random-utility representation of the race ranking. When its scale parameter $\sigma^{\left(r\right)}$ is fixed at one, the Plackett-Luce model provides an analytical solution for the race log likelihood \citep{Glickman2015}:
\begin{equation}
    \label{eq_likelihood_race}
    \ell\left(\boldsymbol{r}_t \, | \, \boldsymbol{f}^{\left(r\right)}, \boldsymbol{\mu}_t^{\left(r\right)}\right) = \sum_{u=1}^{U_t-1} \ddot{\mu}_{u,t}^{\left(r\right)} - \log \sum_{u'=u}^{U_t} \exp\left(\ddot{\mu}_{u',t}^{\left(r\right)}\right),
\end{equation}
where
\begin{equation}
    \ddot{\mu}_{u,t}^{\left(r\right)} = \sum_{d=1}^D \mu_{d,t}^{\left(r\right)} \, \boldsymbol{1}\left(r_{d,t} = U_t + 1 - u\right),
\end{equation}
with $\mu_{d,t}^{\left(r\right)}$ set to some arbitrary quantity if $f_{d,t}^{\left(r\right)}$ equals zero. Recall that a lower performance value $\mu_{d,t}^{\left(r\right)}$ corresponds to a better outcome. Hence, for a given race, equation~\eqref{eq_likelihood_race} factors the race likelihood into the probability that the driver finishing last is outperformed by all others, the probability that the driver finishing second-to-last is outperformed by everyone except the driver finishing last, and so on.

Fixing the Gumbel scale at one normalizes the scale of the latent race performances. Consequently, the coefficients $\beta_1$ and $\beta_2$ determine the magnitude of the shared ability and grid-position contributions relative to the residual variation implied by the ranking model. We refer to this unexplained outcome-specific variation as the \emph{chance component}. This term does not denote pure luck; it collects variation in the observed outcomes that is not explained by the modeled abilities and observed grid position.


The race likelihood in equation~\eqref{eq_likelihood_race} identifies relative rather than absolute latent race performances. Adding the same arbitrary quantity to all race performance variables $\mu_{d,t}^{\left(r\right)}$ leads to the same likelihood. The ranking likelihood therefore does not determine the location of the latent performances. The joint modeling of qualifying times and race rankings provides additional information on the shared driver and constructor states. Because the continuous qualifying equation depends on the same abilities as the race equation, the joint-outcome likelihood restricts common shifts in the shared latent components that could not be determined from the ranking likelihood alone. The fixed Gumbel scale and the restrictions introduced in this section complete the location and scale normalization required for the ranking component.


The additive decomposition of the performance predictors creates an additional identification problem. Adding an arbitrary quantity to all driver abilities and subtracting the same quantity from all constructor abilities leaves both qualifying and race performance predictors unchanged. The joint likelihood therefore cannot distinguish between these parameter values without further restrictions. This difficulty is additional to the identification requirements arising from the dynamic states and partial ranking likelihood. The proposed model must identify not only evolving latent performance, but also its decomposition into two separately evolving states.

We address the above-mentioned invariance by imposing sum-to-zero restrictions on the driver and constructor abilities at every Grand Prix, as well as on their initial abilities:
\begin{equation}
    \sum_{d=1}^D a_{d,t} = \sum_{k=1}^K c_{k,t}
    = \sum_{d=1}^D a_{d,0} + \sum_{k=1}^K c_{k,0} = 0.
\end{equation}
Similar linear constraints are used to identify ability parameters in dynamic ranking models through prior distributions. In particular, we introduce negative correlations in the variance-covariance matrices of the driver and constructor innovations \citep{Glickman2015}:
\begin{align}
    \label{eq_prior_innovations_drv}
    &\boldsymbol{\epsilon}_t^{\left(a\right)} = \pi\left(\sigma^{\left(a\right)} \boldsymbol{\breve{\epsilon}}_t^{\left(a\right)}\right); \quad \boldsymbol{\breve{\epsilon}}_t^{\left(a\right)} \sim \mathcal{N}\left(\boldsymbol{0}, \frac{1}{1 - 1 / V_t} \left(\boldsymbol{I}_{V_t} - \frac{1}{V_t} \boldsymbol{J}_{V_t}\right)\right),\\
    \label{eq_prior_innovations_ctr}
    &\boldsymbol{\epsilon}_t^{\left(c\right)} = \pi\left(\sigma^{\left(c\right)} \boldsymbol{\breve{\epsilon}}_t^{\left(c\right)}\right); \quad \boldsymbol{\breve{\epsilon}}_t^{\left(c\right)} \sim \mathcal{N}\left(\boldsymbol{0}, \frac{1}{1 - 1 / W_t} \left(\boldsymbol{I}_{W_t} - \frac{1}{W_t} \boldsymbol{J}_{W_t}\right)\right),
\end{align}
where $\pi\left(\cdot\right)$ assigns the unscaled innovations to the respective drivers or constructors who participate at Grand Prix $t$. $V_t$ and $W_t$ are the numbers of participating drivers and constructors at Grand Prix $t$, and $\boldsymbol{I}$ and $\boldsymbol{J}$ denote the identity matrix and matrix of ones. The matrices $\boldsymbol{I}_{V_t} - \frac{1}{V_t} \boldsymbol{J}_{V_t}$ and $\boldsymbol{I}_{W_t} - \frac{1}{W_t} \boldsymbol{J}_{W_t}$ introduce the negative innovation correlations required to preserve the zero sums, while the divisions to $1 - 1 / V_t$ and $1 - 1 / W_t$ enforce standard Normal marginal variances. The parameters $\sigma^{\left(a\right)}$ and $\sigma^{\left(c\right)}$ scale the driver and constructor innovations.

The same construction is implemented for the initial abilities:
\begin{align}
    \label{eq_prior_init-abilities_drv}
    &\boldsymbol{a}_0 = \varsigma^{\left(a\right)} \breve{\boldsymbol{a}}_0; \quad \breve{\boldsymbol{a}}_0 \sim \mathcal{N}\left(\boldsymbol{0}, \frac{1}{1 - 1 / D} \, \left(\boldsymbol{I}_D - \frac{1}{D}\boldsymbol{J}_D \right)\right),\\
    \label{eq_prior_init-abilities_ctr}
    &\boldsymbol{c}_0 = \varsigma^{\left(c\right)} \breve{\boldsymbol{c}}_0; \quad \breve{\boldsymbol{c}}_0 \sim \mathcal{N}\left(\boldsymbol{0}, \frac{1}{1 - 1 / K} \, \left(\boldsymbol{I}_K - \frac{1}{K}\boldsymbol{J}_K \right)\right).
\end{align}


The sum-to-zero restrictions determine the relative locations of the driver and constructor states, but the observed driver--constructor assignments must also contain sufficient information to separate the two components. Drivers contracted to the same constructor share the same constructor ability at a given Grand Prix, while retaining different driver abilities. Therefore, differences between teammates are informative about the driver component. When a driver changes constructor, the same driver state becomes associated with a different constructor state, helping to distinguish persistent driver performance from constructor performance over time.

Separating driver and constructor abilities requires variation in the observed driver--constructor assignments. In particular, some drivers must compete for different constructors over the sample, creating the variation needed to distinguish changes in driver ability from changes in constructor ability. Such switches need not occur frequently or between every pair of constructors. This role of mobility in identifying two-way effects is closely related to the worker-firm literature \citep{AbowdKramarzMargolis1999,BonhommeEtAl2023,ChengHoSchorfheide2025}.




The sum-to-zero restrictions are also natural in the context of Formula One. Both standardized qualifying times and race rankings contain information about relative performance. Joint shifts in all driver abilities or all constructor abilities therefore have no direct interpretation in the observed outcomes. Under the Grand Prix-specific centered parameterization, a driver ability of zero corresponds to the average relative driver ability among the competitors represented at time $t$, and a constructor ability of zero corresponds to the corresponding average relative constructor ability. The abilities should therefore be interpreted as deviations from the relevant Grand Prix-specific averages rather than as absolute measures of performance.

The grid variable $g_{d,t}$ is the driver's starting position demeaned within race $t$. Demeaning places grid position on a relative scale, accounts for fluctuations in the number of drivers on the grid, and makes its interpretation compatible with the centered driver and constructor abilities. A value below zero indicates a starting position better than the average grid position at that race.
Similarly, the zero-mean random walks prevent arbitrary shifts in the average driver and constructor abilities over time. This common relative scale facilitates interpretation of the states and allows the contributions of driver and constructor abilities to be compared using posterior summaries.


The joint prior for the model in equations~\eqref{eq_model_core_dgp_quali-time}--\eqref{eq:model_core_dgp-ability-init-ctr} is given by:
\begin{equation}
    \label{eq_prior}
    \begin{split}
        p\left(\boldsymbol{\breve{E}}^{\left(a\right)}, \boldsymbol{\breve{E}}^{\left(c\right)}, \breve{\boldsymbol{a}}_0, \breve{\boldsymbol{c}}_0, \sigma^{\left(a\right)}, \sigma^{\left(c\right)}, \varsigma^{\left(a\right)}, \varsigma^{\left(c\right)}, \sigma^{\left(q\right)}, \boldsymbol{\breve{\beta}}\right).
    \end{split}
\end{equation}
The prior distributions for the unscaled ability innovations $\boldsymbol{\breve{E}}^{\left(a\right)}$ and $\boldsymbol{\breve{E}}^{\left(c\right)}$, as well as for the unscaled initial abilities $\breve{\boldsymbol{a}}_0$ and $\breve{\boldsymbol{c}}_0$, have already been introduced as part of the identification restrictions.

For the standard deviations of the driver and constructor ability innovations and initial abilities, we use the standardization of the qualifying times to determine an appropriate prior scale:
\begin{equation}
    \label{eq_prior_ability-standard-deviations}
    \sigma^{\left(a\right)}, \sigma^{\left(c\right)}, \varsigma^{\left(a\right)}, \varsigma^{\left(c\right)} \sim  \mathcal{N}_{\left(0,\infty\right)}\left(0,1\right).
\end{equation}
These truncated standard Normal priors shrink the state innovation and initial-state scales toward zero while allowing substantial variation relative to the standardized qualifying outcome. Expected performance is therefore primarily determined by the latent abilities, while fluctuations around that expected performance are represented by the outcome-specific residual components.

The priors assign the same marginal distributions to the corresponding driver and constructor scale parameters. They are therefore symmetric at the component-scale level. This symmetry avoids imposing an explicit prior ordering between driver and constructor variability.

Next, we propose a standard Normal prior on the qualifying standard deviation:
\begin{equation}
    \sigma^{\left(q\right)} \sim \mathcal{N}_{\left(0, \infty\right)}\left(0,1\right).
\end{equation}
This prior allows the continuous qualifying outcomes to fluctuate around their expected values on the standardized outcome scale. The parameter $\sigma^{\left(q\right)}$ measures residual qualifying variation not explained by the shared driver and constructor abilities.

Lastly, the race performance coefficients $\beta_1$ and $\beta_2$ share the same truncated Normal prior:
\begin{equation}
    \boldsymbol{\beta} = \kappa \boldsymbol{\breve{\beta}}, \quad \boldsymbol{\breve{\beta}} \sim  \mathcal{N}_{\left(0,\infty\right)}\left(\boldsymbol{0},\boldsymbol{I}_2\right),
\end{equation}
where the choice of $\kappa$ defines the tightness of the distribution. The truncation ensures positive effects of the combined abilities and grid positions on the latent race performances under the sign convention used in the model. Assigning the same marginal prior to both coefficients treats the shared ability component and demeaned grid position symmetrically at the coefficient level.
Since the Gumbel scale is fixed at one, the magnitudes of $\beta_1$ and $\beta_2$ determine the contribution of the systematic race predictor relative to the residual variation implied by the ranking likelihood. Setting $\kappa$ to five permits the race ranking to be strongly associated with either the shared ability component or grid position while retaining regularization of extreme coefficient values.

The priors in~\eqref{eq_prior} are weakly informative in the sense that they provide mild regularization on scales defined relative to the standardized qualifying outcomes. Still, their influence should be assessed relative to the likelihood and the induced prior predictive behavior of the model.


The model is estimated using Stan's No-U-Turn Sampler (NUTS) \citep{hoffman2014no}, a Hamiltonian Monte Carlo algorithm suitable for high-dimensional and differentiable posterior distributions. The model combines a non-Gaussian ranking likelihood, dynamic latent states, and correlated Gaussian constructions that impose the identification restrictions. NUTS permits joint posterior inference for the latent abilities, state innovation scales, observation parameters, and race coefficients while retaining the full posterior uncertainty.

We use NUTS rather than a standard Kalman filter because the partial ranking observation equation is not linear and Gaussian. Variational inference could provide a computationally faster approximation, but common variational families may underestimate posterior uncertainty or inadequately represent complex posterior dependence \citep{blei2017variational,lichter2024variational}. For the present application, we therefore use MCMC inference as the primary estimation method.

\section{Empirical Application: The Formula One Hybrid Era}
\label{sec_case-study}


We apply the model introduced in Sections~\ref{sec_model}--\ref{sec_identification_priors} to qualifying and race outcomes from the Formula One hybrid era, spanning eight seasons from 2014 to 2021. This period was the most recent fully completed regulatory era at the outset of the project. An era refers to a period during which the principal technical regulations remain broadly comparable across seasons, for example with respect to engine specifications \citep{Bhambwani2023}. The sample contains 160 Grand Prix weekends across 30 circuits, involving 51 drivers and 12 constructors after accounting for constructor rebrandings.

\subsection{Data and identifying variation}

The observed driver--constructor assignments provide the variation required to separate the two dynamic ability components. At a given Grand Prix, drivers contracted to the same constructor share a constructor state but have separate driver states. Across Grand Prix weekends, driver movements between constructors associate the same driver with different constructor states. These teammate comparisons and assignment changes jointly provide information about the driver--constructor decomposition.

\begin{figure}[htbp]
\centering
\includegraphics[width=1\linewidth]{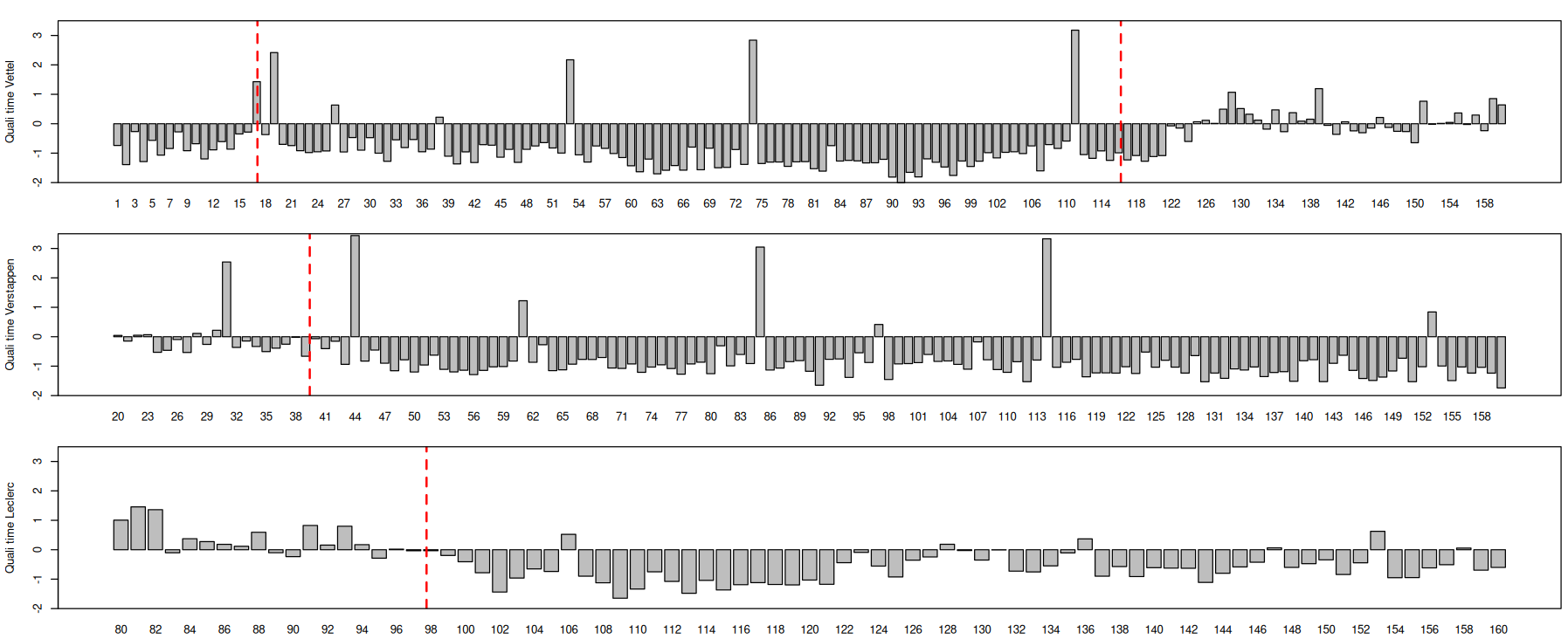}
\caption{Qualifying times for Vettel, Verstappen, and Leclerc}
\label{fig_descriptive-stats_quali-times}
\vspace{5pt}
\begin{minipage}{\textwidth}
\small The y-axes show standardized fastest qualifying lap times and the x-axes show Grand Prix weekends. The vertical red lines indicate constructor changes. Lower values correspond to better qualifying outcomes.
\end{minipage}
\end{figure}

\begin{figure}[htbp]
\centering
\includegraphics[width=1\linewidth]{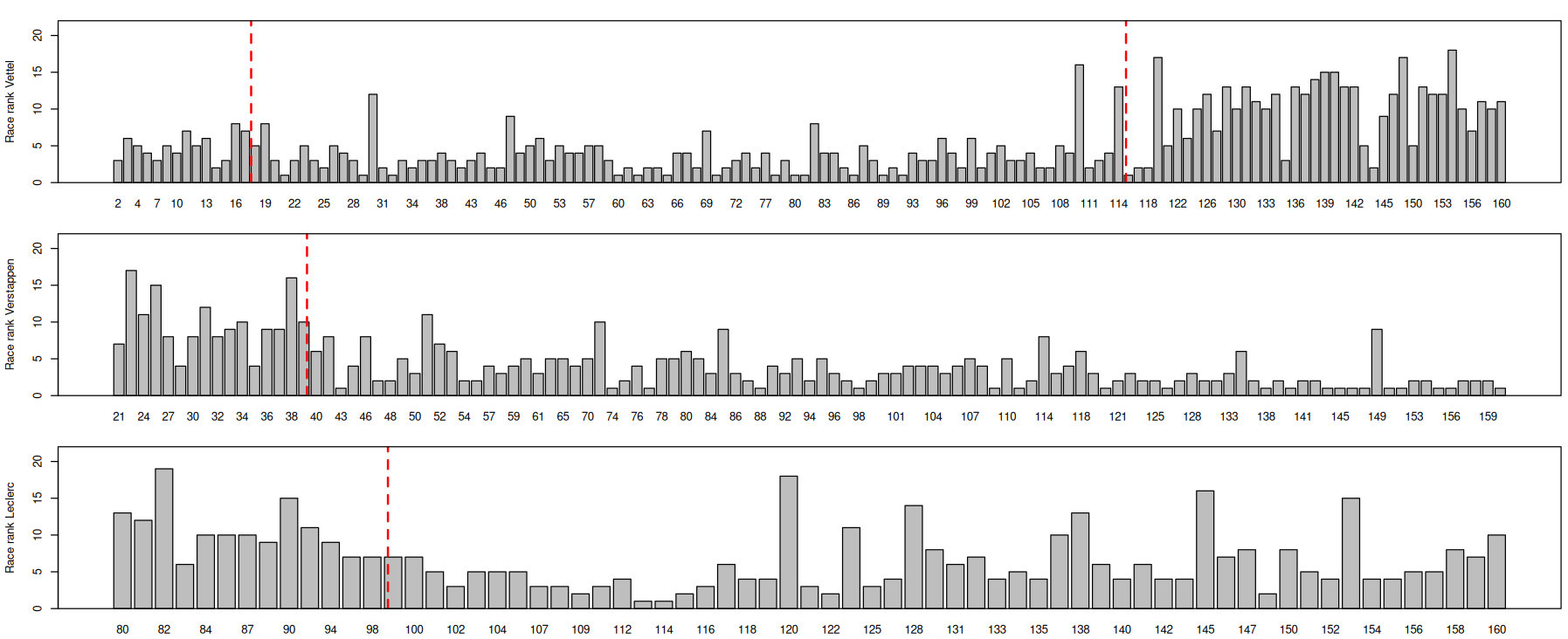}
\caption{Race ranks for Vettel, Verstappen, and Leclerc}
\label{fig_descriptive-stats_race-ranks}
\vspace{5pt}
\begin{minipage}{\textwidth}
\small The y-axes show official race ranks and the x-axes show Grand Prix weekends. The vertical red lines indicate constructor changes. Lower values correspond to better race outcomes.
\end{minipage}
\end{figure}

Figures~\ref{fig_descriptive-stats_quali-times} and~\ref{fig_descriptive-stats_race-ranks} present the observed qualifying times and race ranks for Vettel, Verstappen, and Leclerc. All three drivers changed constructors during the hybrid era, as indicated by the vertical lines.

The observed qualifying times and race ranks vary substantially over the Grand Prix weekends. Some changes occur gradually while a driver remains with the same constructor, while other changes coincide with a constructor change. For Vettel, both outcomes generally deteriorate after his move from Ferrari to Aston Martin. Verstappen's outcomes improve sharply after moving from Toro Rosso to Red Bull and continue to improve during his time with Red Bull. Leclerc's qualifying times and race ranks also improve following his move from Sauber to Ferrari.

These patterns illustrate the empirical difficulty addressed by the model. Abrupt changes around driver movements may contain information about constructor ability, while gradual changes within an assignment may reflect changes in the driver, constructor, or both. These descriptive results are not sufficient to determine the relevant decomposition, particularly because driver and constructor abilities evolve simultaneously.


\subsection{Posterior computation}

We estimate the model using the NUTS algorithm. The results are based on four MCMC chains, each consisting of 2000 warm-up iterations and 2000 sampling iterations. Convergence is assessed using chain-specific and parameter-specific diagnostics recommended for Stan.

At the chain level, we monitor divergent transitions, maximum-treedepth exceedances, and the Bayesian fraction of missing information (BFMI). Divergent transitions indicate that the sampler has not explored part of the posterior density, while maximum-treedepth exceedances indicate potential computational inefficiency. BFMI values below 0.3 indicate potentially poor exploration of the marginal energy distribution.

At the parameter level, we evaluate the potential scale-reduction factor, $\widehat{R}$, and the bulk and tail effective sample sizes. We require $\widehat{R}$ values below 1.01 and effective sample sizes of at least 400. The estimation does not lead to divergent transitions or maximum-treedepth exceedances, all BFMI values exceed 0.3, and the parameter-specific diagnostics satisfy the above-mentioned criteria.

\subsection{Dynamic driver--constructor decomposition}

\begin{figure}[htbp]
\centering
\begin{subfigure}{1\textwidth}\centering
\includegraphics[width=\linewidth]{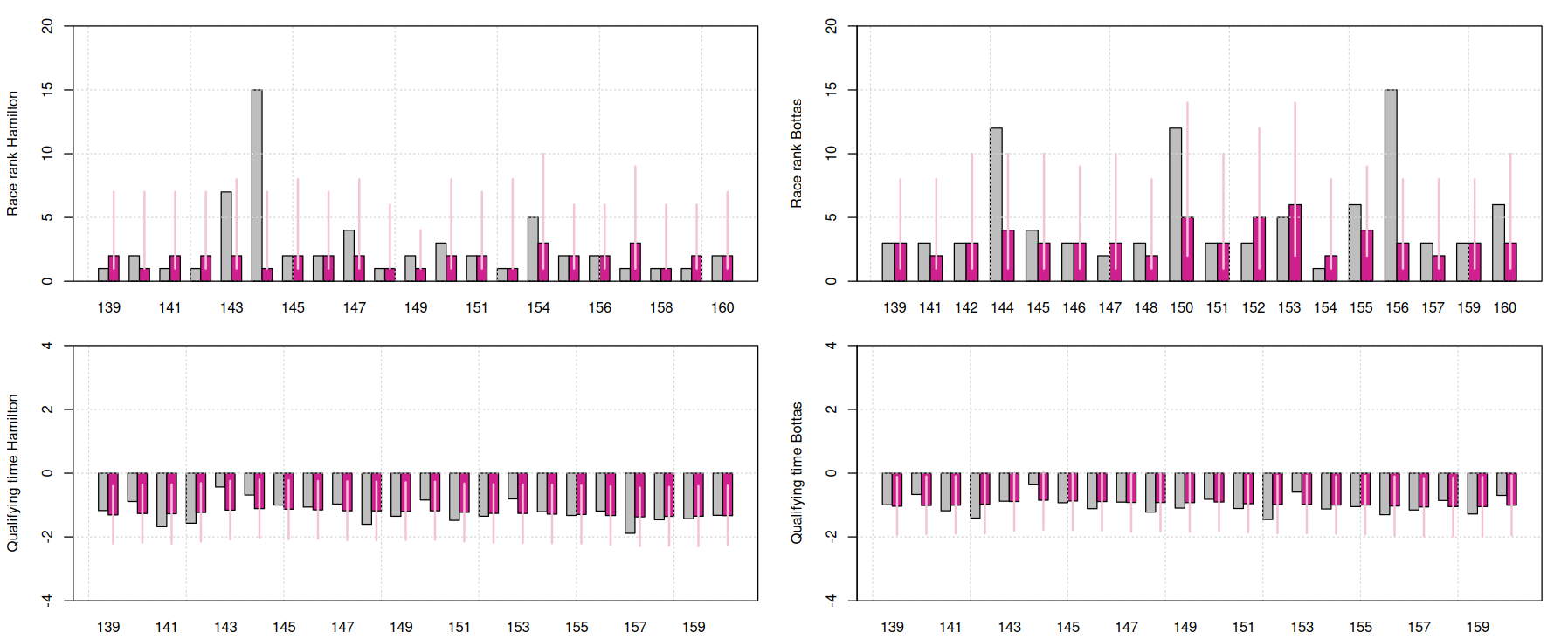}
\caption{Race ranks and qualifying times}
\vspace{10pt}
\end{subfigure}
\begin{subfigure}{1\textwidth}\centering
\includegraphics[width=\linewidth]{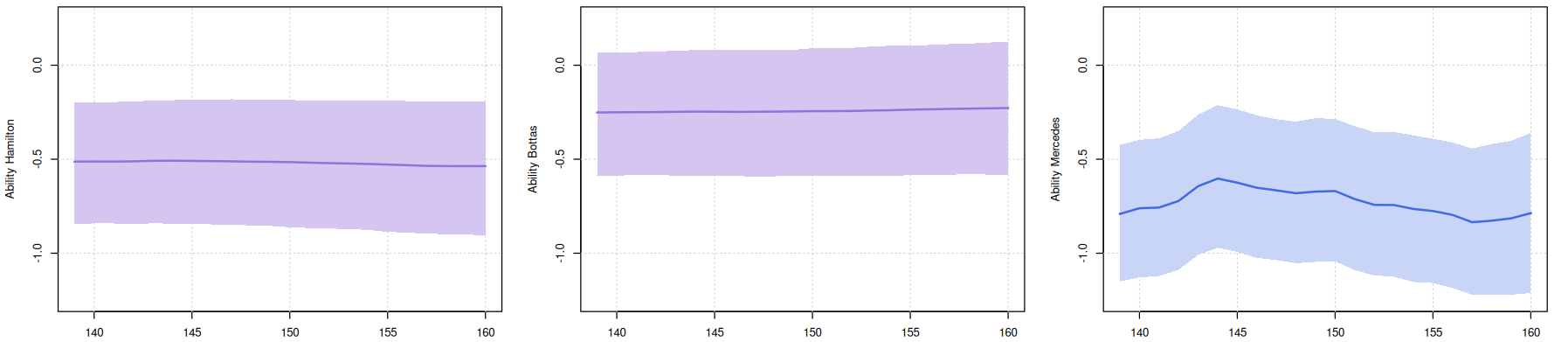}
\caption{Driver and constructor abilities}
\end{subfigure}
\caption{Dynamic decomposition of Mercedes performance in the 2021 season}
\label{fig_disentanglement-merc}
\vspace{5pt}
\begin{minipage}{\textwidth}
\small The x-axes show Grand Prix weekends. In panel a), the first row presents observed race ranks in grey and posterior medians in pink, including middle 90 percent credible intervals, for races finished by Hamilton and Bottas. The second row presents observed standardized qualifying times in grey and posterior means in pink, including middle 90 percent credible intervals. Panel b) presents the posterior mean abilities of Hamilton, Bottas, and Mercedes, including middle 90 percent credible intervals. Lower values correspond to better ability.
\end{minipage}
\end{figure}

The principal empirical output of the model is the decomposition of shared Formula One performance into separately evolving driver and constructor abilities. Figure~\ref{fig_disentanglement-merc} illustrates this decomposition for Mercedes during the 2021 season. Hamilton and Bottas were both contracted to Mercedes, so their qualifying and race outcomes depend on separate driver states but the same constructor state.

The top panel of Figure~\ref{fig_disentanglement-merc} compares the observed outcomes with their posterior summaries. Hamilton generally achieves better qualifying times and race ranks than Bottas, and this difference is also reflected in the fitted outcomes. The credible intervals remain relatively wide for individual Grand Prix weekends, particularly for the race rankings.
The bottom panel presents the corresponding decomposition. Hamilton's posterior mean ability is generally better than Bottas's, although the credible intervals indicate substantial uncertainty about the magnitude of this difference. Both driver abilities evolve relatively gradually during the season. Hamilton's ability improves slightly, whereas Bottas's ability deteriorates slightly.

The Mercedes constructor ability exhibits larger within-season movements than either driver trajectory. Because Hamilton and Bottas share this state, common movements in their qualifying and race outcomes provide information about the constructor component, while persistent differences between them provide information about the separate driver components. The model therefore does not attribute every change in an individual driver's outcomes to that driver alone.

For much of the season, the Mercedes constructor component also has a larger absolute posterior mean than either driver component. Within this example, the constructor therefore makes a substantial contribution to the shared performance predictor. We note that these comparisons are conditional on the model, the relative zero-centered scale, and the posterior uncertainty surrounding all three states.

\subsection{Abilities across the hybrid era}

\begin{figure}[htbp]
\centering
\includegraphics[width=1\linewidth]{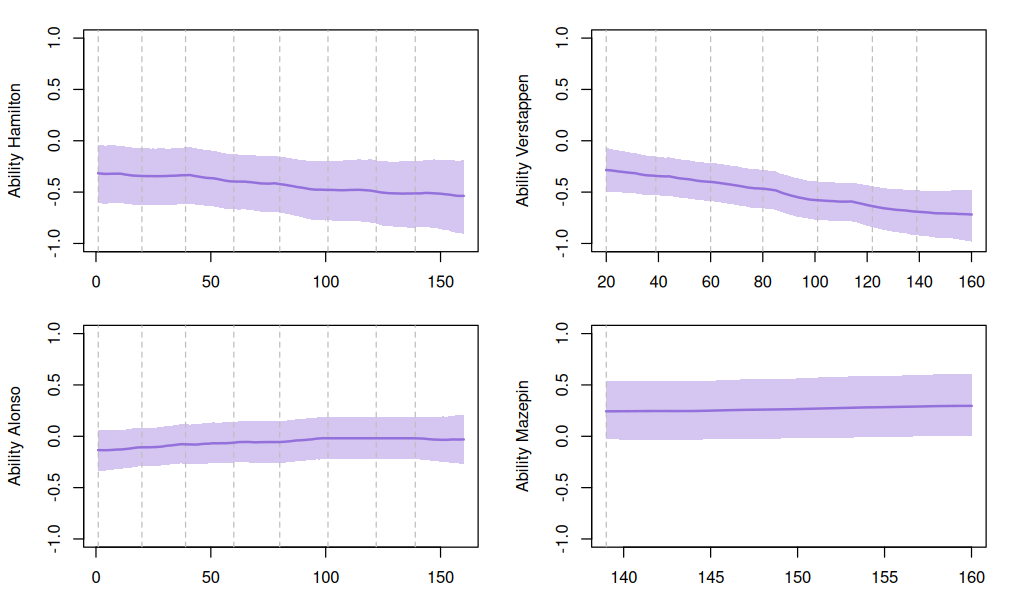}
\caption{Driver abilities during the hybrid era}
\label{fig_drviver-abilities}
\vspace{5pt}
\begin{minipage}{\textwidth}
\small The y-axes show posterior mean driver abilities, including middle 90 percent credible intervals. Lower values correspond to better ability. The x-axes show Grand Prix weekends from the first to the last Grand Prix entered by each driver during the hybrid era. The vertical grey lines mark the first Grand Prix of each season.
\end{minipage}
\end{figure}

\begin{figure}[htbp]
\centering
\includegraphics[width=\linewidth]{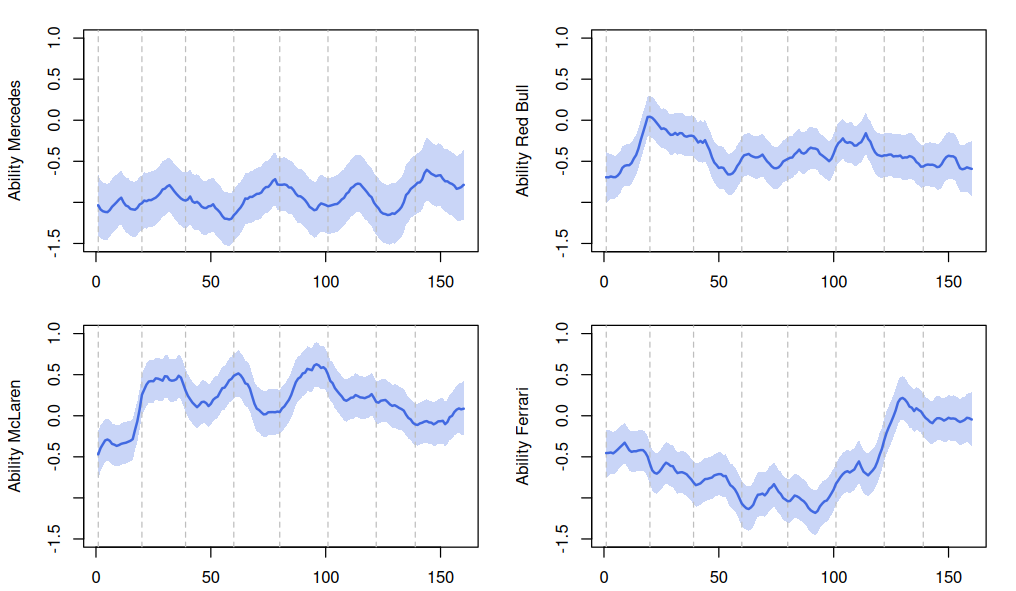}
\caption{Constructor abilities during the hybrid era}
\label{fig_constructor-abilities}
\vspace{5pt}
\begin{minipage}{\textwidth}
\small The y-axes show posterior mean constructor abilities, including middle 90 percent credible intervals. Lower values correspond to better ability. The x-axes show Grand Prix weekends from the first to the last Grand Prix entered by each constructor during the hybrid era. The vertical grey lines mark the first Grand Prix of each season.
\end{minipage}
\end{figure}

Figures~\ref{fig_drviver-abilities} and~\ref{fig_constructor-abilities} extend the decomposition to selected drivers and constructors across the full hybrid era. The selected trajectories indicate that driver abilities are generally more stable than constructor abilities. This pattern is consistent with the Mercedes example but is not imposed by the symmetric component-scale priors. The model nevertheless allows drivers to improve or deteriorate gradually. Verstappen provides the clearest example of sustained improvement during the hybrid era, while Hamilton's posterior mean ability also improves more moderately over time.

Constructor abilities exhibit more pronounced changes within and across seasons. Mercedes begins the hybrid era with a particularly strong constructor state, while Red Bull gradually improves its performance. Ferrari's constructor ability deteriorates during the middle of the sample before becoming more stable toward the end of the period. The figures therefore illustrate two different forms of dynamic variation. Driver trajectories tend to move gradually, whereas constructor trajectories contain larger and sometimes more abrupt changes. This does not by itself establish that constructors always contribute more to performance. Relative contribution depends on the posterior levels, dispersion, and evolution of both components for each driver--constructor assignment. For several combinations, however, the assigned constructor component has a larger absolute contribution to the shared predictor than the corresponding driver component.


\subsection{Posterior predictive assessment}

We next assess whether the proposed model can reproduce the two heterogeneous outcome types. For continuous qualifying outcomes, uncertainty can be summarized using posterior means and credible intervals. Such pointwise summaries are less informative for race rankings because the ranks within each race are jointly constrained: assigning one driver to a particular rank necessarily changes the ranks available to the remaining drivers. We therefore use complete replicated outcome sequences drawn from the posterior predictive distribution.

\begin{figure}[htbp]
\centering
\includegraphics[width=1\linewidth]{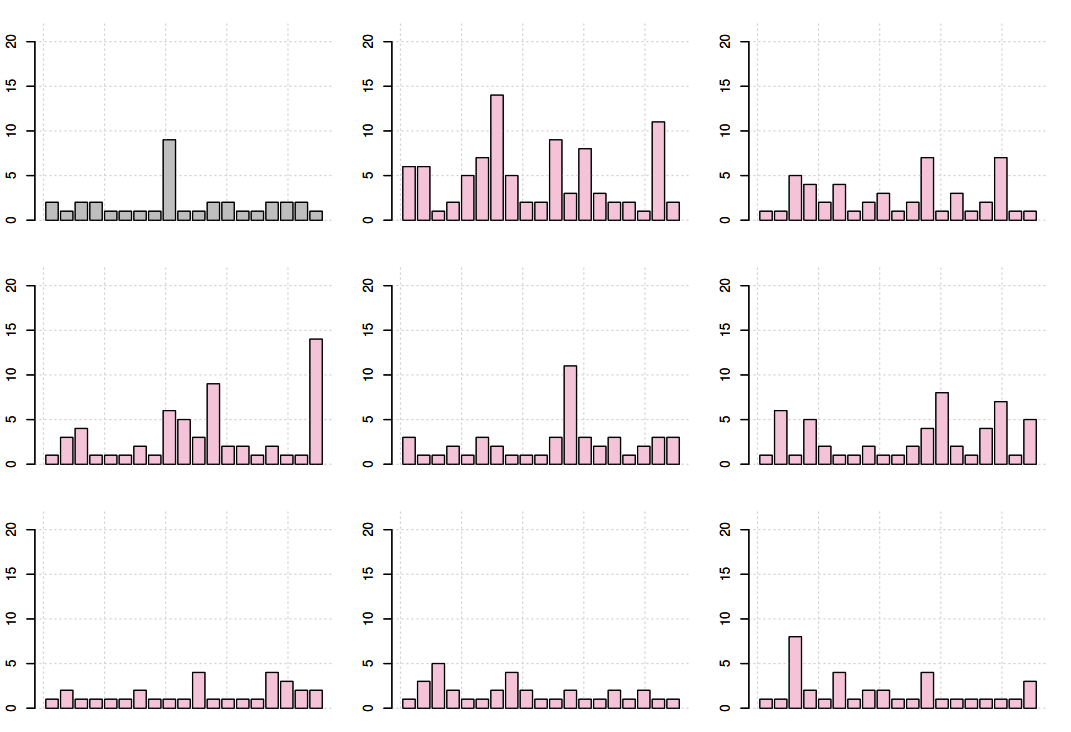}
\caption{Grand Prix-level race posterior predictive assessment for Verstappen in the 2021 Red Bull}
\label{fig_ppc_race_verstappen}
\vspace{5pt}
\begin{minipage}{\textwidth}
\small The top-left panel shows the observed race ranks. Each remaining panel shows one complete replicated sequence drawn from the posterior predictive distribution after thinning to eight draws. Each bar represents Verstappen's race rank at the corresponding Grand Prix.
\end{minipage}
\end{figure}

Figure~\ref{fig_ppc_race_verstappen} compares Verstappen's observed race ranks with eight replicated race sequences. In this figure, each panel represents one realization of the ranking process over the season. Variation across the panels reflects posterior predictive variability, which combines uncertainty about the model parameters and latent states with outcome-specific residual variation. The posterior predictive draws generally characterize the Verstappen--Red Bull pairing as a leading competitor, with most replicated outcomes concentrated near the top of the field. The observed sequence is nevertheless more dominant: Verstappen ranks within the top two in all but one of the included races, whereas several replicated sequences contain more outcomes around fifth place or below.

The differences among the replicated panels show that the model implies substantial predictive variability for individual race outcomes. They also reveal a systematic feature of the fit: although the model places Verstappen and Red Bull among the leading competitors, it tends to understate the frequency with which this pairing occupies the very top ranks. This leading driver--constructor pairing between Verstappen and Red Bull during the 2021 season enables the assessment of the model near the top of the race-ranking distribution. Corresponding checks for Alonso in the 2021 Alpine and Mazepin in the 2021 Haas are presented in Appendix~\ref{sec_appendix}, allowing the model's behavior to be compared in a different part of the race-ranking distribution.

\begin{figure}[htbp]
\centering
\includegraphics[width=1\linewidth]{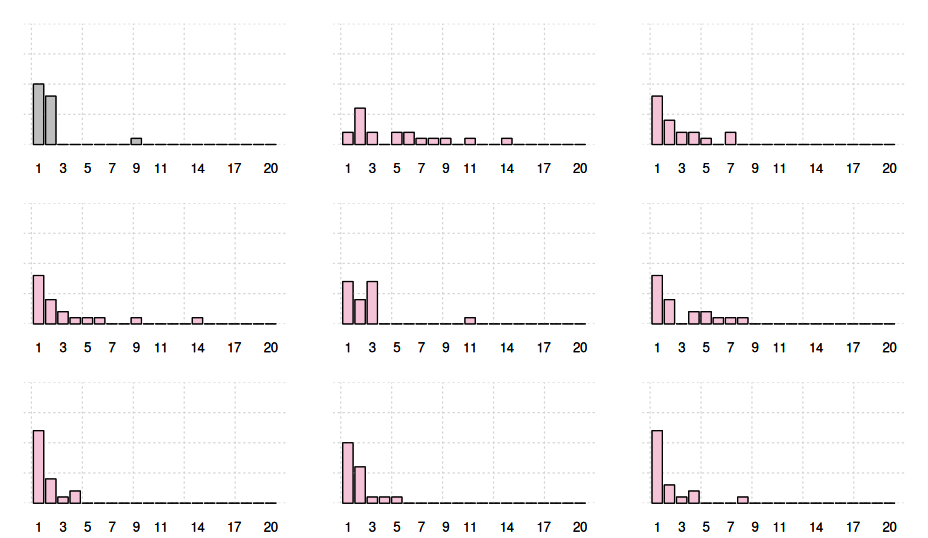}
\caption{Aggregated race posterior predictive assessment for Verstappen in the 2021 Red Bull}
\label{fig_ppc_race-aggr_verstappen}
\vspace{5pt}
\begin{minipage}{\textwidth}
\small The top-left histogram shows the observed distribution of race ranks. Each remaining histogram shows one replicated seasonal distribution drawn from the posterior predictive distribution after thinning to eight draws. Each bar represents the number of races in which Verstappen obtained the corresponding rank.
\end{minipage}
\end{figure}

Figure~\ref{fig_ppc_race-aggr_verstappen} aggregates the ranks across the 2021 season, and presents the Grand Prix-level comparison. This comparison provides additional information since a replicated rank does not need to match the observed rank at the same race, even if the obtained seasonal ranking distribution is similar. The aggregated display confirms the conclusion from the Grand Prix-level assessment. The replicated distributions place substantial probability on leading ranks, but they contain more outcomes around fifth place or below than the observed distribution. The model therefore reproduces Verstappen's general position near the front of the field but does not fully reproduce the concentration of his observed outcomes in the top two positions.

\begin{figure}[htbp]
\centering
\includegraphics[width=1\linewidth]{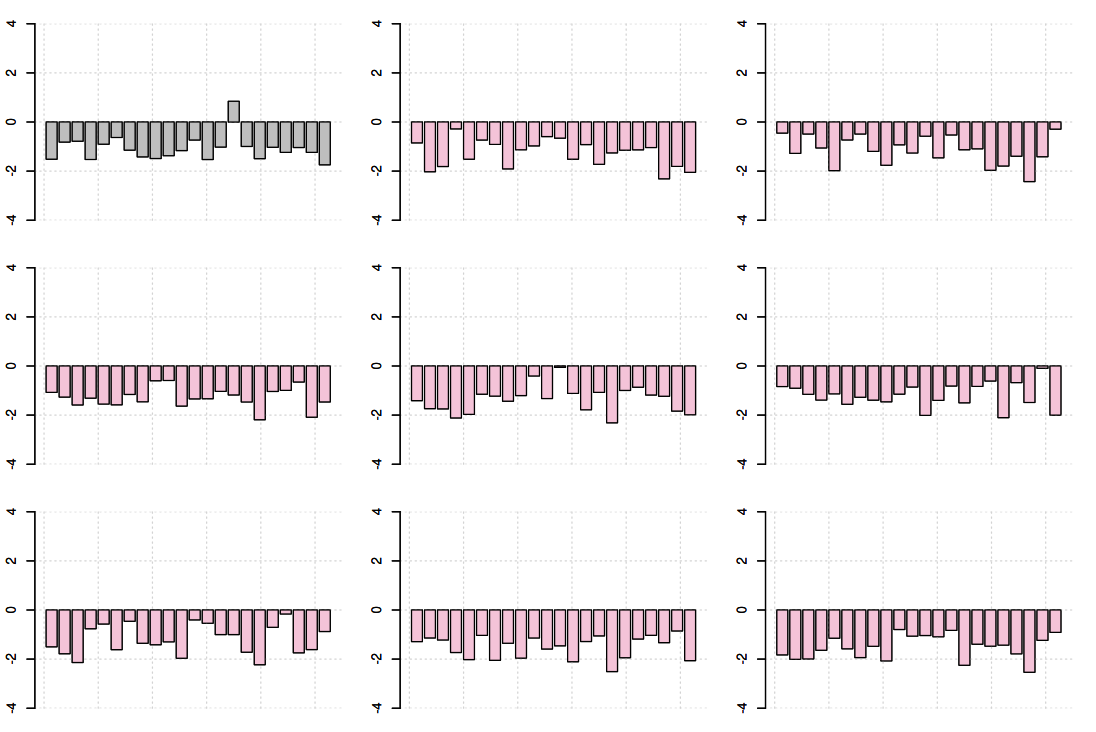}
\caption{Grand Prix-level qualifying posterior predictive assessment for Verstappen in the 2021 Red Bull}
\label{fig_ppc_quali_verstappen}
\vspace{5pt}
\begin{minipage}{\textwidth}
\small The top-left panel shows the observed standardized qualifying times. Each remaining panel shows one complete replicated sequence drawn from the posterior predictive distribution after thinning to eight draws. Each bar represents Verstappen's qualifying time at the corresponding Grand Prix.
\end{minipage}
\end{figure}

\begin{figure}[htbp]
\centering
\includegraphics[width=1\linewidth]{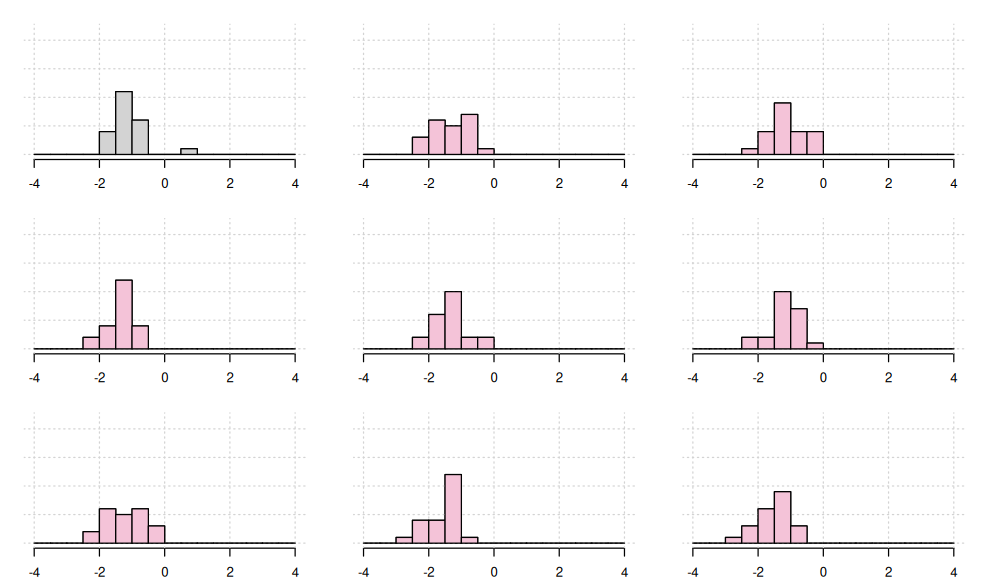}
\caption{Aggregated qualifying posterior predictive assessment for Verstappen in the 2021 Red Bull}
\label{fig_ppc_quali-aggr_verstappen}
\vspace{5pt}
\begin{minipage}{\textwidth}
\small The top-left histogram shows the observed distribution of standardized qualifying times. Each remaining histogram shows one replicated seasonal distribution drawn from the posterior predictive distribution after thinning to eight draws.
\end{minipage}
\end{figure}

Figures~\ref{fig_ppc_quali_verstappen} and~\ref{fig_ppc_quali-aggr_verstappen} present the corresponding checks for qualifying times. The replicated qualifying sequences exhibit substantial Grand Prix-level variation, but the observed qualifying outcomes do not stand out systematically from the posterior predictive draws. The agreement is particularly clear in the aggregated display, where the replicated seasonal distributions are similar to the observed distribution.

For the selected pairing, the model therefore reproduces the continuous qualifying outcome more closely than the upper tail of the race-ranking outcome. The difference may be related to the distinct observation models: qualifying times are modeled using a Normal distribution, whereas the race outcome is generated through a Gumbel random-utility representation and the resulting Plackett-Luce likelihood. The results reported in Appendix~\ref{sec_appendix} indicate whether the same discrepancy is present for middle- and lower-performing pairings.

\subsection{Uncertainty in the latent abilities}

\begin{figure}[htbp]
\centering
\includegraphics[width=0.7\linewidth]{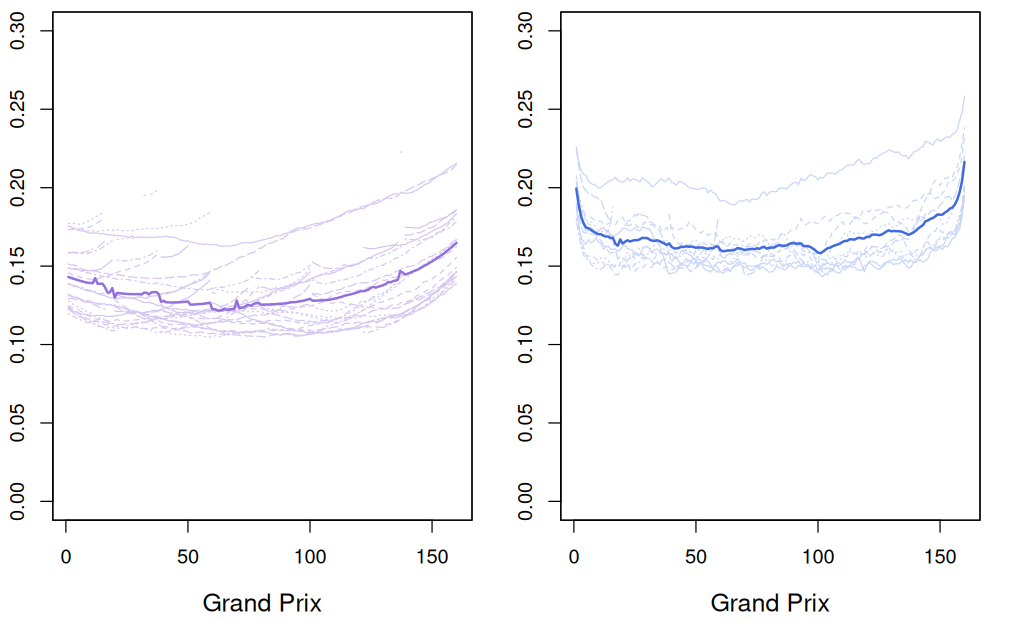}
\caption{Posterior uncertainty in driver and constructor abilities}
\label{fig_post-ability-certainties}
\vspace{5pt}
\begin{minipage}{\textwidth}
\small Each non-bold line shows the posterior standard deviation of an individual driver's ability on the left or constructor's ability on the right. The bold lines represent the average posterior standard deviation across participating drivers or constructors at each Grand Prix. Nonparticipating drivers and constructors are excluded from the corresponding averages.
\end{minipage}
\end{figure}

Figure~\ref{fig_post-ability-certainties} summarizes uncertainty in the estimated driver and constructor states using their posterior standard deviations over time. Posterior uncertainty for both driver and constructor abilities is generally higher near the beginning and end of the sample and lower in the middle. This pattern follows from smoothing in a state-space model: states in the interior of the sample are informed by observations both before and after the corresponding Grand Prix, whereas states near the boundaries have less information on one side. The posterior standard deviations of constructor abilities are generally higher than those of driver abilities. This indicates that the constructor states are estimated less precisely under the current model and data. The reported posterior uncertainty partly reflects variation in constructor ability, but it also depends on the effective number of observations, the assignment structure, and uncertainty in the remaining model parameters.

The application illustrates the three principal features of the proposed model. The continuous qualifying outcome and partial race ranking jointly inform shared latent states; these states are decomposed into separately evolving driver and constructor abilities; and the resulting trajectories permit their contributions to be compared over time. The posterior predictive assessment indicates that the model reproduces qualifying outcomes reasonably well in the examined cases, while the race-ranking component has greater difficulty reproducing the sustained dominance of a leading driver--constructor pairing.

\section{Conclusion}
\label{sec_conclusion}

We propose a Bayesian state-space model that decomposes observed Formula One outcomes into dynamic driver and constructor abilities. The model jointly analyzes two heterogeneous outcome types: standardized qualifying times, modeled as continuous outcomes, and race results, modeled as partial rankings. Both outcomes depend on the same dynamic driver and constructor states, while retaining outcome-specific sampling distributions.

The proposed model provides a dynamic decomposition of shared latent performance. Driver and constructor abilities are represented by two separately evolving states, and the observed contractual assignments determine which constructor ability enters each driver's performance predictor. Identification of this decomposition combines information from the joint-outcome likelihood, sum-to-zero restrictions implemented through the prior specification, and variation in driver--constructor assignments over time. The assignment structure is particularly important because teammate comparisons and driver movements between constructors provide information for separating the two latent components.

We apply the model to qualifying and race outcomes from the Formula One hybrid era between 2014 and 2021. The posterior estimates indicate that driver abilities are generally more stable over time, although some drivers exhibit sustained improvement or deterioration. Constructor abilities show larger movements within and across seasons and, for several driver--constructor combinations, make a larger contribution to the shared performance predictor. The decomposition also separates persistent differences between teammates from changes that are shared by drivers contracted to the same constructor. The posterior predictive assessment reproduces the qualifying outcomes relatively well in the examined cases, while the race-ranking model generates greater variability and does not fully reproduce the sustained dominance of a leading driver--constructor pairing.

The model combines elements from mixed-outcome latent-variable models, dynamic ranking models, and assignment-based two-way decompositions within a common Bayesian state-space framework. Although the model is developed for Formula One, the same structure may be useful in other settings where heterogeneous longitudinal outcomes depend jointly on separately evolving individual and group effects and where group memberships change over time. Examples include organizational outcomes and other team-based competitions in which the contributions of individuals and the groups to which they are assigned are both of substantive interest.

\bibliographystyle{chicago}
\bibliography{references}

\clearpage
\appendix

\counterwithin{figure}{section}
\counterwithin{table}{section}

\section{Posterior predictive checks}
\label{sec_appendix}
In this appendix, we present additional posterior predictive checks for Alonso in the 2021 Alpine (medium competitor) and for Mazepin in the 2021 Haas (bottom competitor):


\begin{figure}[htbp]
    \centering
    \includegraphics[width=1\linewidth]{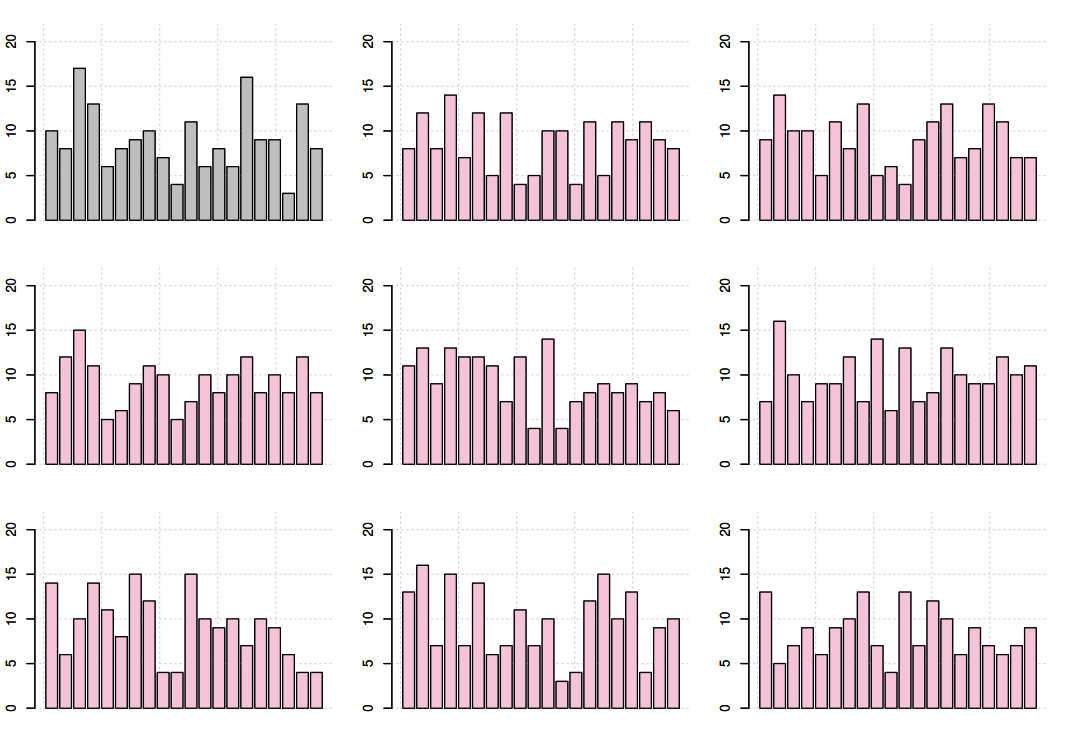}
    \caption{Race PPC for Alonso (Alpine) in 2021 season}
    \label{fig_ppc_race_alonso}
\end{figure}
\begin{figure}[htbp]
    \centering
    \includegraphics[width=1\linewidth]{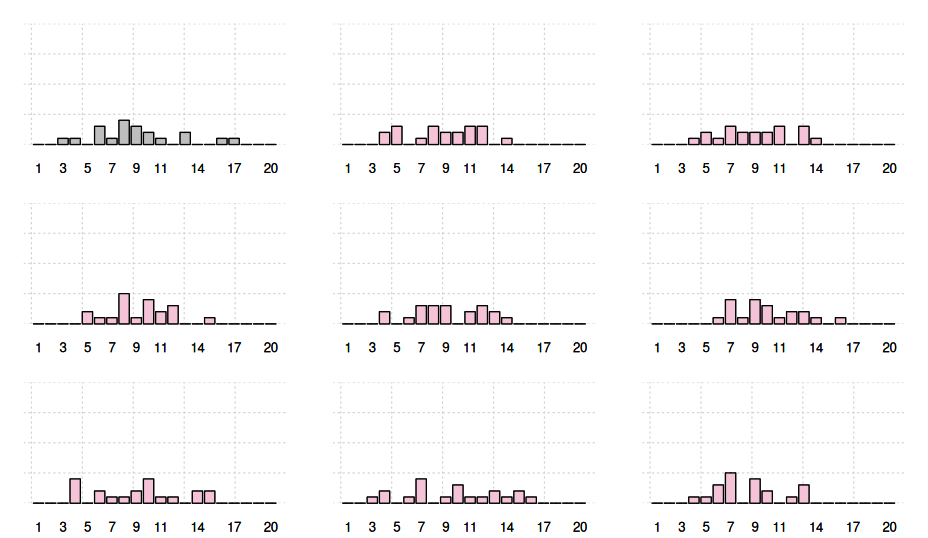}
    \caption{Aggregated Race PPC for Alonso (Alpine) in 2021 season}
    \label{fig_ppc_race-aggr_alonso}
\end{figure}
\begin{figure}[htbp]
    \centering
    \includegraphics[width=1\linewidth]{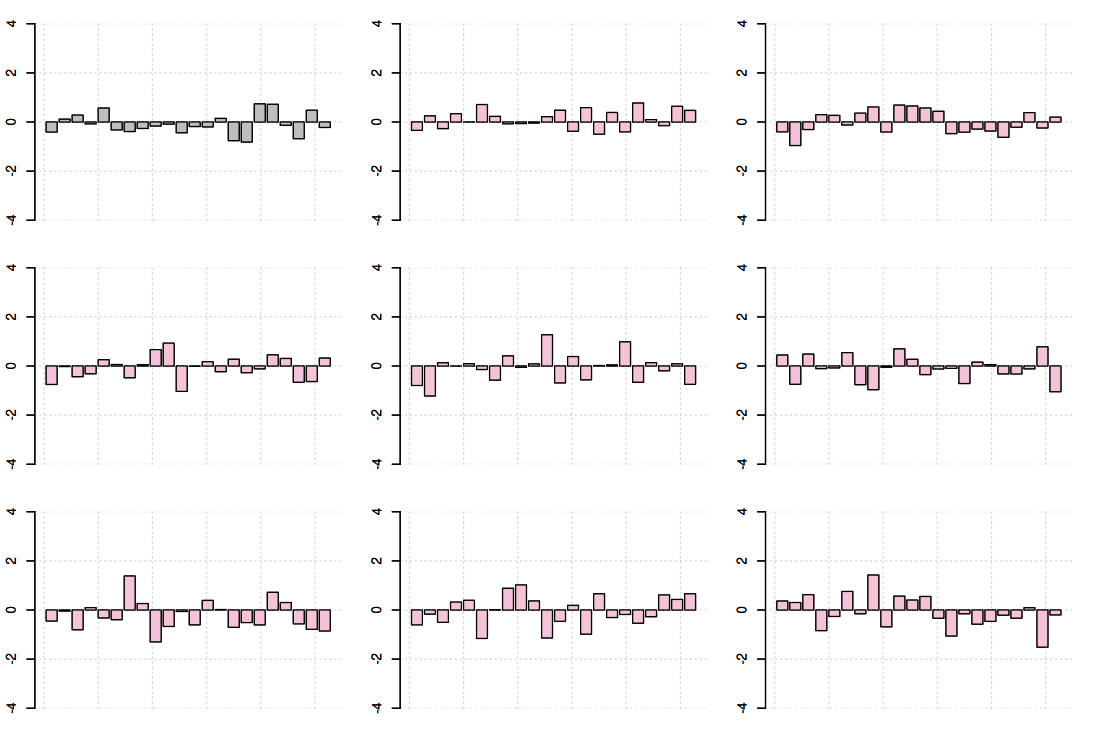}
    \caption{Qualifying PPC for Alonso (Alpine) in 2021 season}
    \label{fig_ppc_quali_alonso}
\end{figure}
\begin{figure}[htbp]
    \centering
    \includegraphics[width=1\linewidth]{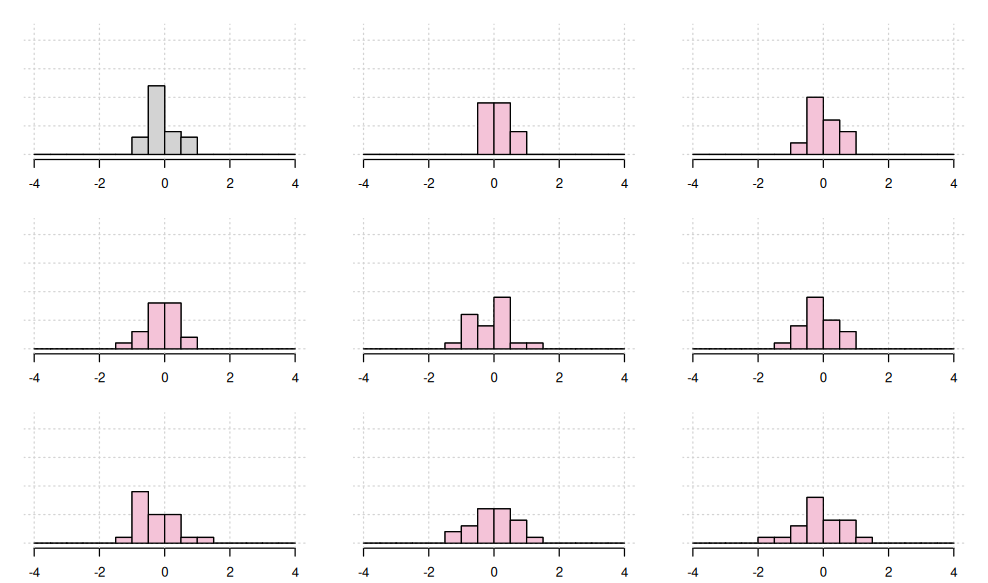}
    \caption{Aggregated Qualifying PPC for Alonso (Alpine) in 2021 season}
    \label{fig_ppc_quali-aggr_alonso}
\end{figure}
\begin{figure}[htbp]
    \centering
    \includegraphics[width=1\linewidth]{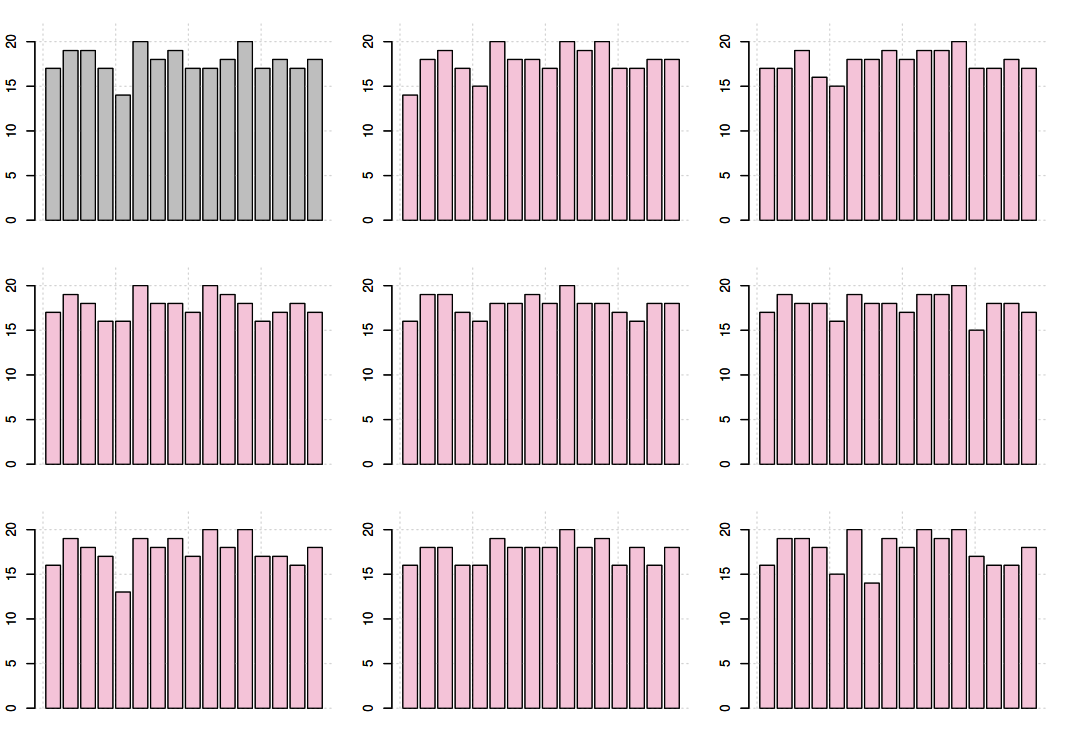}
    \caption{Race PPC for Mazepin (Haas) in 2021 season}
    \label{fig_ppc_race_mazepin}
\end{figure}
\begin{figure}[htbp]
    \centering
    \includegraphics[width=1\linewidth]{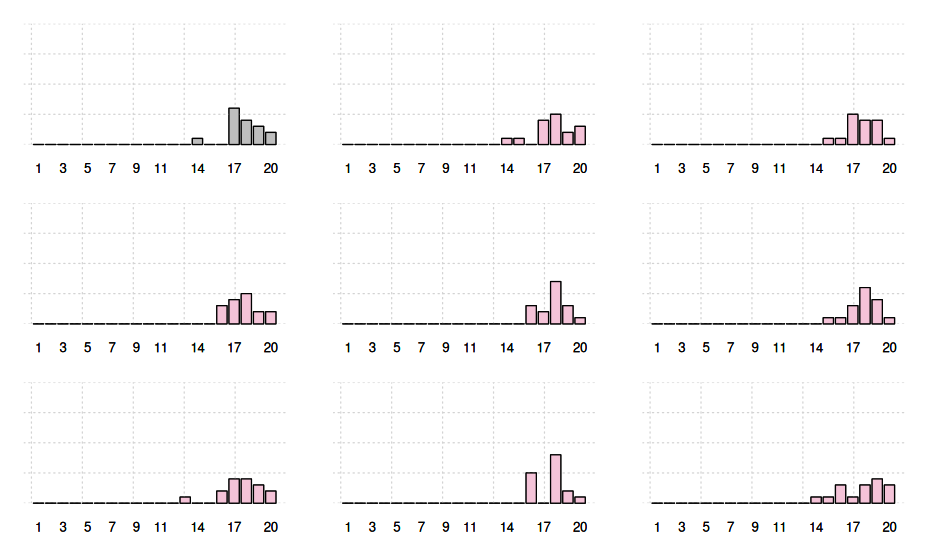}
    \caption{Aggregated Race PPC for Mazepin (Haas) in 2021 season}
    \label{fig_ppc_race-aggr_mazepin}
\end{figure}
\begin{figure}[htbp]
    \centering
    \includegraphics[width=1\linewidth]{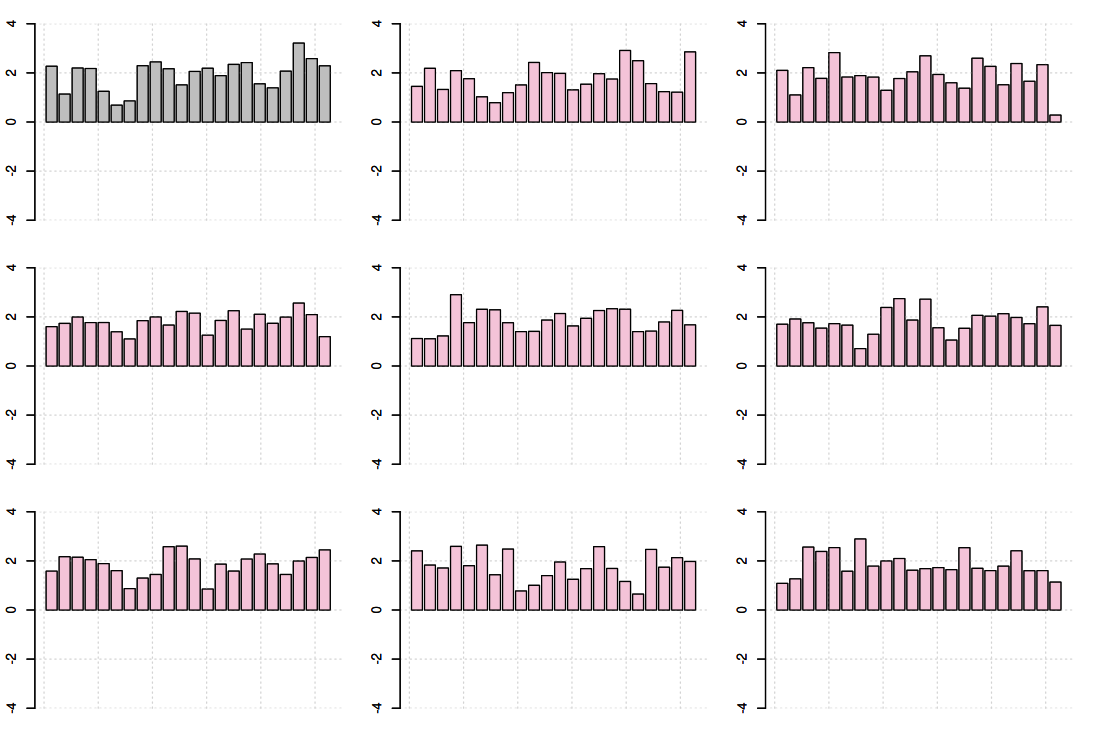}
    \caption{Qualifying PPC for Mazepin (Haas) in 2021 season}
    \label{fig_ppc_quali_mazepin}
\end{figure}
\begin{figure}[htbp]
    \centering
    \includegraphics[width=1\linewidth]{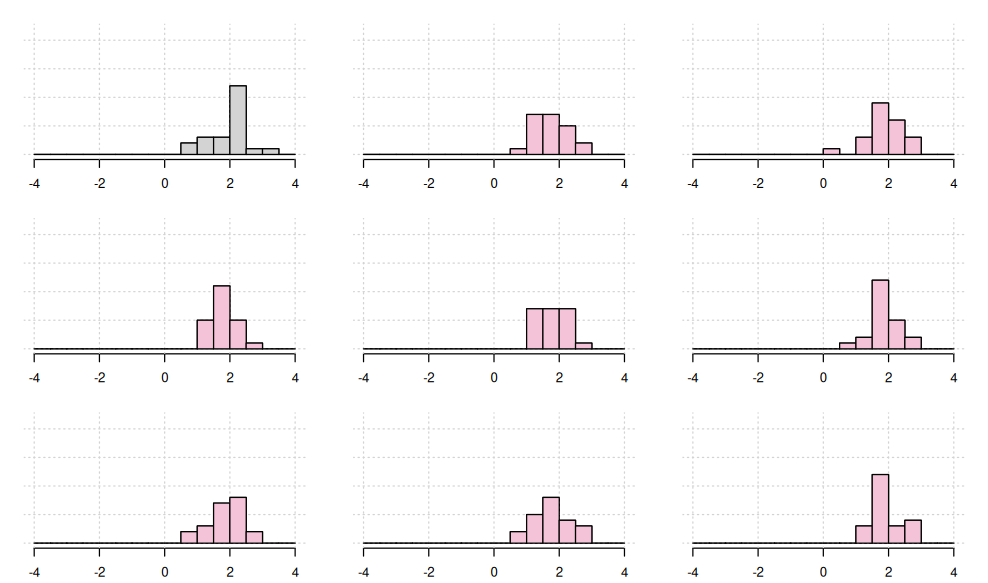}
    \caption{Aggregated Qualifying PPC for Mazepin (Haas) in 2021 season}
    \label{fig_ppc_quali-aggr_mazepin}
\end{figure}


\end{document}